\documentclass[twocolumn]{autart}    

\usepackage{graphicx} 
\usepackage{amsmath,amssymb, bm}
\usepackage{mathtools}
\usepackage{algorithm}
\usepackage{algpseudocode}
\newtheorem{lemma}{Lemma}
\newtheorem{theorem}{Theorem}
\newtheorem{assumption}{Assumption}
\newtheorem{remark}{Remark}
\newtheorem{problem}{Problem}
\newtheorem{subproblem}{Problem}[problem]

\begin{document}

\begin{frontmatter}

\title{Dynamic Question Design for Efficient Estimation of\\ Aggregate Human Preferences} 


\author[Keio]{Kazuyoshi Fukuda}\ead{kazu150207@keio.jp},    
\author[Keio]{Masaki Inoue}\ead{minoue.z6@keio.jp},               

\address[Keio]{Department of Applied Physics and Physico-Informatics, Keio University, 3-14-1 Hiyoshi, Kohoku-ku, Yokohama, Kanagawa, Japan}  


\begin{abstract}                          
This paper addresses the problem of efficiently estimating aggregate human preferences by dynamically adapting questions based on respondents’ answers.
To this end, we formulate and address two sub-problems: (1) preference estimation and (2) question design.
First, regarding preference estimation, we model respondents' preferences and estimate them using Bayesian estimation, employing a particle filter as a computationally efficient approximation. The main theoretical contribution to this sub-problem is to analyze the preference estimation error using an information-theoretic approach, deriving a theoretical lower bound for the error.
Second, regarding question design, we formulate the design problem as an Expected Information Gain (EIG) maximization problem and employ an $\epsilon$-greedy strategy to solve the problem in a computationally efficient way. We theoretically analyze the search efficiency of the approach, demonstrating that it achieves higher efficiency than a random search. 
Finally, we verify the effectiveness of the proposed method through numerical simulations.
\end{abstract}

\end{frontmatter}

\section{Introduction}\label{intro}
Estimating people's needs through questionnaires is important in both the business and public sector. In business, questionnaires are useful tools for understanding customer needs. By using questionnaires to identify the specifications, designs, and prices that customers want, companies can provide products that better meet customer demand. This can reduce unnecessary production and improve profitability. In public administration, questionnaires help governments understand citizens' needs and develop more livable communities efficiently with limited financial resources.

Question design in a questionnaire plays a central role in accurately and efficiently estimating respondents' needs. Studies in economics have also reported that the content of questions has a significant impact on survey results \cite{Pref,Consume}. Let us consider a simple example. Suppose that each respondent selects one product from a set of $n$ products. In this paper, the probability of selecting each product is referred to as the ``preference''. We consider the problem of estimating these preferences from questionnaire responses. A straightforward question for this purpose is: “If you were to choose one of these $n$ products, which would you select?” Although this question is simple and direct, presenting a large number of alternatives imposes a substantial cognitive burden on respondents. Consequently, respondents may be more likely to decline to answer or provide responses that do not accurately reflect their true preferences. Indeed, previous studies have shown that presenting respondents with too many alternatives increases their cognitive burden\cite{Schwartz2004}.

One approach to reducing the burden is to present respondents with a subset of the available alternatives and ask them to select one. Although each response to such a question may provide less information, the reduced number of alternatives is expected to facilitate decision-making and increase the response rate. Other possible question formats include partitioning the alternatives into several groups and asking respondents to select one group, or presenting three alternatives and asking them to select two. Thus, questionnaires can be designed using a wide variety of question formats. However, it remains unclear which format enables the most efficient estimation of respondents’ preferences.


In this paper, we address two problems: estimating respondents' needs from the obtained responses and 
dynamically adapting questions based on respondents’ answers.
We consider aggregate respondents rather than a single individual. Specifically, we focus on choice settings in which each respondent is presented with multiple alternatives and asked to select one. A representative example is a purchasing decision in which a consumer selects one product from several available options. Our primary objective is to estimate the aggregate respondents preference for each alternative, defined as the probability that the alternative is selected.

As discussed above, questionnaire-based preference estimation involves two interrelated processes. The first process sequentially updates the preference estimates based on the responses collected at each time step. The second dynamically determines which question should be presented next to improve estimation efficiency. For the first process, established statistical methods, such as Bayesian inference, can be employed. Bayesian inference yields a posterior distribution over the preferences rather than a single point estimate, thereby explicitly quantifying the uncertainty associated with the estimates.
However, Bayesian estimation generally incurs a high computational cost. Thus, various computational methods have been proposed to address this issue \cite{mcmc_population,mcmc_bayesian,mcmc_single,variational,variational_blind,particle}. 
For dynamic question design, methods from optimal experimental design provide a natural foundation \cite{choice-based,active,opt-based}. Expected information gain (EIG) \cite{EIG,multi} is a widely used criterion in Bayesian optimal experimental design and has been applied in diverse fields, including bioinformatics \cite{bio}, neuroscience \cite{nerve}, psychology \cite{psy}, and criminal investigation \cite{eye}. EIG quantifies the expected reduction in uncertainty about the unknown preferences resulting from a candidate question. Nevertheless, EIG-based question design can incur a substantial computational cost because the EIG must be evaluated for each candidate question and each possible response. This computational burden increases rapidly with the number of alternatives and candidate questions, making straightforward EIG-based approaches unsuitable for large-scale sequential questionnaire design.

Optimal experimental design is closely related to the field of systems and control. The design of maximally informative experiments for parameter estimation has a long history in system identification; see the classical survey \cite{mehra1974} and the more recent overviews \cite{hjalmarsson2005,pronzato2008}. In particular,  \cite{pronzato2008} highlights the intimate connection between optimal experimental design and control problems, and \cite{jansson2005,bombois2006} formulate input design as optimization problems under practical constraints. The sequential structure of our problem, in which questions are updated dynamically on the basis of the current posterior, is analogous to adaptive input design and dual control \cite{feldbaum1960,heirung2017}, where excitation for learning and exploitation of the current estimate must be balanced.

To address the challenges outlined above, we adopt the following approaches. First, we employ a particle filter \cite{particle} to reduce the computational burden of sequential preference estimation. In addition to enabling efficient recursive updates, particle filters can accommodate flexible, potentially non-Gaussian prior and posterior distributions. Second, we dynamically optimize both the number and the combination of alternatives presented in each question. This design is intended to reduce the burden on respondents while improving the informativeness of their responses. To further reduce the computational cost of question design, we introduce an optimization method based on an $\epsilon$-greedy strategy \cite{eps_ant,eps_babbling}. Although an $\epsilon$-greedy strategy is commonly used in multi-armed bandit problems, we apply it to a deterministic combinatorial optimization problem arising in dynamic question design. We further derive a theoretical lower bound on the Bayes risk associated with the proposed preference estimation method. In addition, we analyze the reliability of the $\epsilon$-greedy question-design method by evaluating the probability that it identifies the globally optimal question. Finally, we demonstrate the effectiveness of the proposed framework through numerical simulations.

\section{System Overview}
\begin{figure}[t]
 \centering 
 \includegraphics[width=1.05\linewidth]{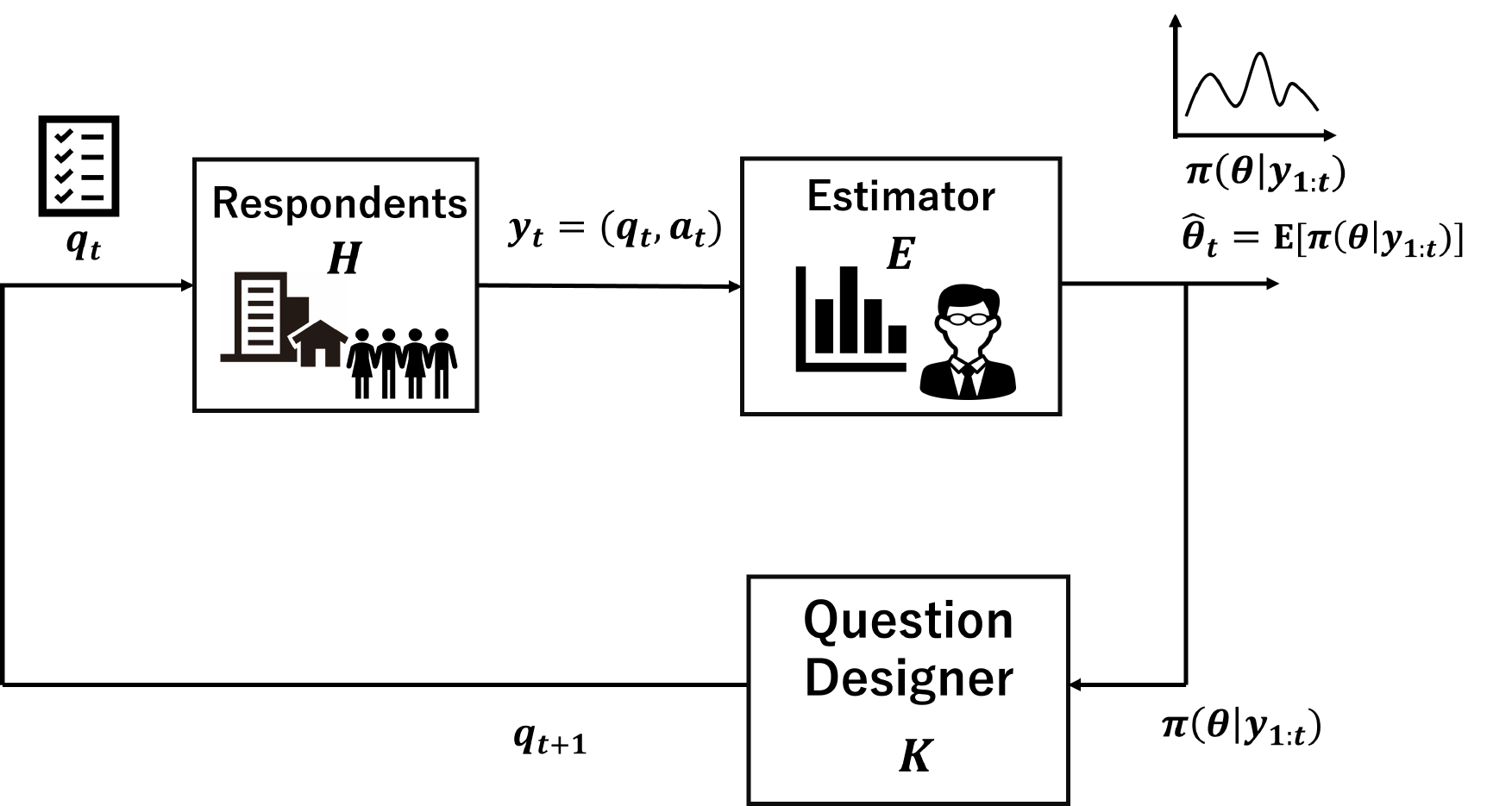} 
 \caption{Overview of the proposed system}
 \label{overview}
\end{figure}

This paper aims to estimate aggregate human preferences, focusing on aggregate respondents preferences rather than individual ones. To improve estimation efficiency, we develop an adaptive question design system that dynamically adjusts the questions presented to respondents. The overall structure of the system proposed in this paper is shown in Fig.~\ref{overview}.
The proposed system consists of three blocks: $H$, which represents aggregate respondents who answer questions; $E$, which estimates the respondents' preferences from the obtained responses; and $K$, which optimizes the next question based on the estimated posterior distribution of the preferences.

The respondents $H$ have true probabilities with which they choose each alternative from among $n$ alternatives. These probabilities are referred to as the true preferences and are denoted by $\theta=(\theta_1,\dots,\theta_n)^T$. Given a presented question $q_t \subseteq \{1,2,\ldots,n\}$, the respondents $H$ make a choice and provide a response $a_t$.

The estimator $E$ is responsible for estimating the true preferences $\theta$ of the respondents $H$. Using the sequence of observations up to time $t$, denoted by $y_{1:t}:=(y_1,\dots,y_{t})$ where $y_t=(q_t,a_t)$, the estimator $E$ obtains the posterior distribution of the respondents' preferences and its expectation as the estimated preference by using Bayesian estimation. 

\begin{figure}[t]
 \centering 
 \includegraphics[width=1.05\linewidth]{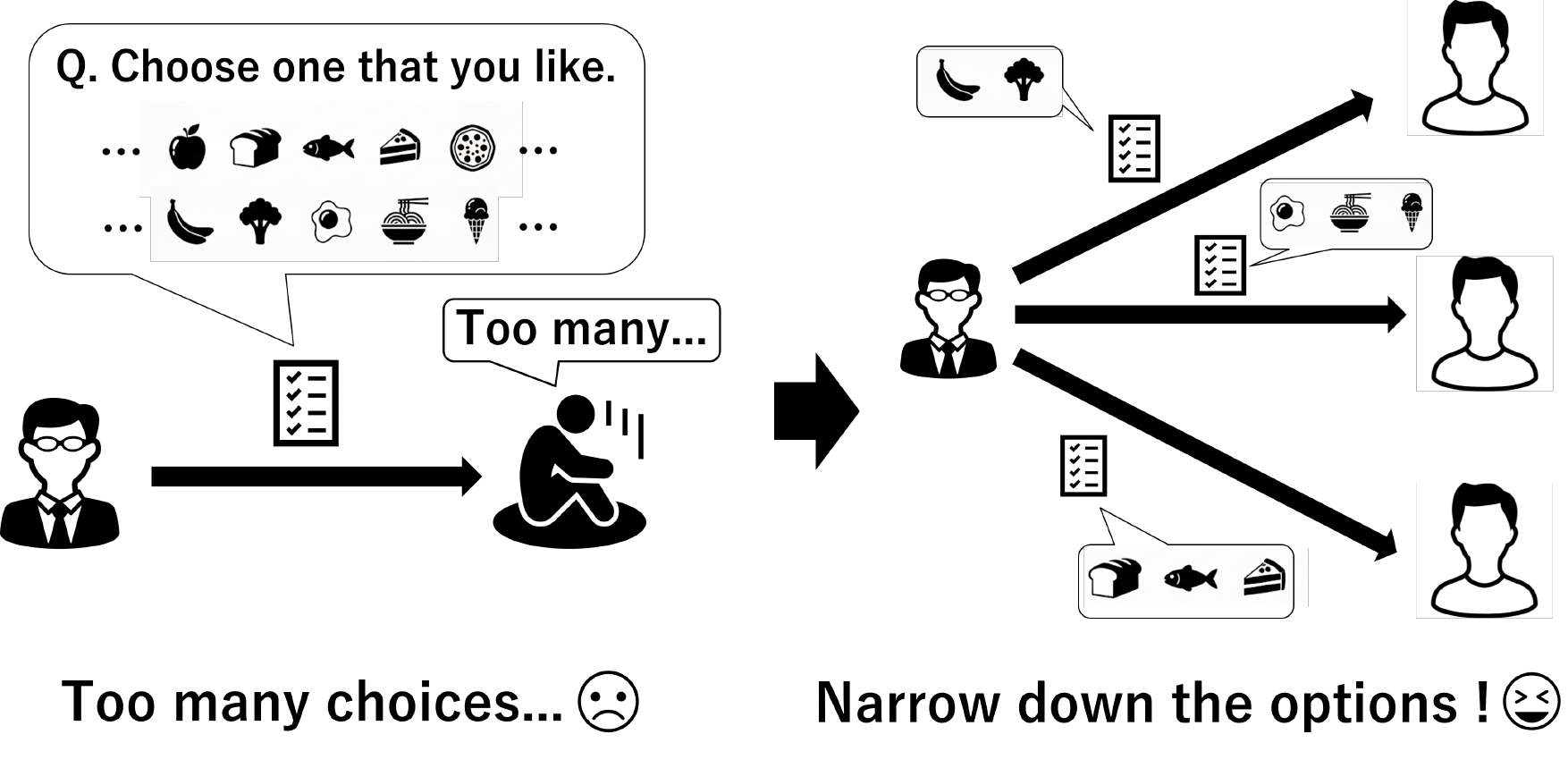} 
 \caption{The conceptual diagram of $q$}
 \label{question_image}
\end{figure}

The question designer $K$ designs a question $q_t$ from the full set of alternatives and presents it to the respondents $H$. $q_t$ is a narrowed-down selection of a few options from a total of $n$ alternatives. For example, letting $q_t =\{1,3,7\}$, the corresponding question asks, ``Which of alternatives 1, 3, and 7 do you prefer?'' Hereafter, with a slight abuse of notation, we identify each question with its associated set of alternatives and denote both by $q_t$. The conceptual diagram of $q_t$ is shown in Fig.~\ref{question_image}.
As shown in Fig.~\ref{question_image}, $q_t$ is a subset of the $n$ alternatives. The question designer $K$ uses the posterior distribution obtained by $E$ to dynamically optimize and design the next question $q_{t+1}$ to estimate the true preferences $\theta$, which is then presented to $H$.

By repeating this procedure composed of question design by $K$, response by $H$, and preference estimation by $E$---the preferences $\theta$ of the respondents $H$ can be estimated efficiently.

The details of the respondents $H$ and the estimator $E$ are described in Section~\ref{esitimator}, and the details of the question designer $K$ are described in Section~\ref{designer}.



\section{Respondent Preference Estimator $E$}\label{esitimator}

In this section, we address the problem of estimating the preferences of respondents $H$.
First, we construct a model of the respondents $H$, characterized by their probabilistic responses $a_t$ to question $q_t$. We then present a method for estimating the preferences of the modeled respondents $H$.

\subsection{Respondents $H$}\label{respondents}

In this subsection, we construct a model of responses to question $q$ designed by the question designer $K$ based on the discrete choice model \cite{logit}.

We prepare the following assumption.
\begin{assumption}
The preference vector $\theta=(\theta_1,\dots,\theta_n)$ assigned to each alternative is time-invariant.   
\end{assumption}
Under Assumption 1, the question designer $K$ selects a subset of alternatives from the $n$ alternatives and constructs a question $q_t$. Then, the probability that the response $a_t$ to question $q_t$ corresponds to alternative $i$ is given by
\begin{equation}\label{logit2}
\Pr(a_t=i \mid q_t,\theta) = \frac{\theta_i}{ \sum_{j\in q_t} \theta_j}.
\end{equation}
Recall that the question $q_t$ represents a set of $|q_t|$ alternatives presented to respondents. Respondents are asked to choose one alternative from this ``set $q_t$''.

\subsection{Respondent Preference Estimator $E$}\label{estimator_sub}

In this subsection, we present a method of estimating the preferences of respondents $H$ by Bayesian estimation.

\subsubsection{Bayesian Preference Estimation}\label{sub_bayes_update}

In general, Bayesian estimation updates the probability distribution $\pi(\theta)$ over the unknown parameter $\theta = (\theta_1,\dots,\theta_n)$ based on observation data $y$ and Bayes' theorem, thereby obtaining a posterior distribution that reflects the observation data.

Let the prior distribution at time $t-1$ and the likelihood function be denoted by $\pi(\theta \mid y_{1:t-1})$ and $p(y_t \mid \theta)$, respectively. Then, according to Bayes' theorem, the posterior distribution $\pi(\theta \mid y_{1:t})$ obtained after observing $y_t$ in addition to $y_{1:t-1}$ is expressed as
\begin{align}\label{bayes_update}
\pi(\theta \mid y_{1:t})
&= \frac{p(y_{t} \mid \theta)\pi(\theta \mid y_{1:t-1})}
{\displaystyle\int p(y_{t} \mid \theta')\pi(\theta'\mid y_{1:t-1})d\theta'} \nonumber\\
&= \frac{p(y_{t} \mid \theta)\pi(\theta\mid y_{1:t-1})}
{p(y_{t}\mid y_{1:t-1})}.
\end{align}

In this paper, we use \eqref{logit2} as the likelihood function for Bayesian estimation. Using the pair $y_t=(q_t, a_t)$ consisting of the question $q_t$ at time $t$ and the corresponding response $a_t$, the likelihood function $p(y_{t} \mid \theta)$ is written as
\begin{equation}\label{likelihood}
p(y_{t} \mid \theta)=p(a_t \mid q_t, \theta).
\end{equation}
$p(a_t \mid q_t, \theta)$ is given by \eqref{logit2}.
From \eqref{logit2}, \eqref{bayes_update} and \eqref{likelihood}, the posterior distribution $\pi(\theta \mid y_{1:t})$ given $\pi(\theta \mid y_{1:t-1})$ is obtained as
\begin{equation}\label{bayes_likelihood}
\pi(\theta \mid y_{1:t})
= \frac{\frac{\theta_{a_t}}{\sum_{i \in q_t} \theta_i}\pi(\theta \mid y_{1:t-1})}{\displaystyle\int\frac{\theta'_{a_t}}{\sum_{i \in q_t} \theta'_i}\pi(\theta'\mid y_{1:t-1})d\theta'}.
\end{equation}

\subsubsection{Approximate Estimation Using a Particle Filter} \label{subsec:ParticleFilter}
Since the integral; $\int\frac{\theta'_{a_t}}{\sum_{i \in q_t} \theta'_i}\pi(\theta'\mid y_{1:t-1})d\theta'$, the Bayesian update in \eqref{bayes_likelihood} is difficult to calculate directly.
Therefore, we employ a particle filter \cite{particle} to approximately perform the Bayesian update in \eqref{bayes_likelihood}. The particle filter represents the posterior distribution $\pi_{t}(\theta \mid y_{1:t})$ by an approximate distribution $\tilde{\pi}_{t}(\theta \mid y_{1:t})$ consisting of a set of particles.
We explain how the Bayesian update described above is performed using a particle filter.

We first sample from the approximate distribution at the previous time step $\tilde{\pi}_{t-1}(\theta \mid y_{1:t-1})$, and obtain $L$ particles $\{\theta_{t-1}^i\}_{i=1}^{L}$.
Then, using the particles $\{\theta_{t-1}^i\}_{i=1}^{L}$, the approximate prior distribution at the current time step $t$, $\tilde{\pi}_{t} (\theta \mid y_{1:t-1})$, is described as follows:
\begin{align}\label{particle_previous}
\tilde{\pi}_{t}(\theta \mid y_{1:t-1}) = \frac{1}{L} \sum_{i=1}^{L} \delta(\theta  - \theta_{t-1}^i).
\end{align}
The approximate posterior distribution $\tilde{\pi}_t (\theta \mid y_{1:t})$ is obtained by assigning weights to each delta function $\delta(\theta  - \theta_{t-1}^i)$ in the approximate prior distribution $\tilde{\pi}_{t} (\theta \mid y_{1:t-1})$ according to the observation $y_t$.
First, the weight $\alpha_t^i$ for each delta function $\delta(\theta  - \theta_{t-1}^i)$ is computed as follows:
\begin{align}\label{weight_alpha}
\alpha_t^i = \frac{p (y_t \mid \theta_{t-1}^i)}{\sum_{j=1}^{L} p (y_t \mid \theta_{t-1}^j)}, \quad \forall i \in \{1, \dots, L\}.
\end{align}
$p (y_t \mid \theta_{t-1}^i)$ is the likelihood function and computed using \eqref{logit2}.
In the above equation, the $L$ weights $\alpha_t^i$ are normalized so that their sum equals $1$.
Using the weights $\alpha_t^i$, the approximate posterior distribution at the current time step $t$, $\tilde{\pi}_t (\theta \mid y_{1:t})$, is computed as follows:
\begin{align}\label{particle_posterior}
\tilde{\pi}_t(\theta \mid y_{1:t}) = \sum_{i=1}^{L} \alpha_t^i  \delta(\theta  - \theta_{t-1}^i).
\end{align}
The particles $\{\theta_{0}^i\}_{i=1}^{L}$ are obtained by sampling from the prior distribution $\pi_0 (\theta)$ at the initial time $t=0$. The prior distribution $\pi_0 (\theta)$ is specified in advance.

We assume that if no response is obtained at time $t$, the approximate posterior distribution is not updated and remains unchanged from time $t-1$\footnote{When no response is obtained at time $t$, $a_t$ records that the question $q_t$ was unanswered.}. We define a binary variable $r(q_t)$, which takes the value 1 if a response is obtained for question $q_t$ and 0 otherwise. Accordingly,
\begin{equation}
\tilde{\pi}_t(\theta \mid y_{1:t})
=
\begin{cases}
\displaystyle
\sum_{i=1}^{L} \alpha_t^i \delta(\theta-\theta_{t-1}^i),
& r(q_t)=1,\\[2mm]
\tilde{\pi}_{t-1}(\theta \mid y_{1:t-1}),
& r(q_t)=0.
\end{cases}
\end{equation}

The preference estimation method proposed in this section is summarized in Algorithm \ref{alg:particle_update}.

\begin{algorithm}[t]
\caption{Particle-Based Bayesian Estimation of Preference Parameters}
\label{alg:particle_update}
\begin{algorithmic}[1]
\Require Prior distribution $\pi_0(\theta)$, number of particles $L$,
questions $\{q_t\}_{t=1}^{T}$
\Ensure Approximate posterior distribution
$\tilde{\pi}_T(\theta \mid y_{1:T})$

\State Sample $\{\theta_0^i\}_{i=1}^{L}$ from $\pi_0(\theta)$
\State Set $\alpha_0^i \gets 1/L$ for all $i\in\{1,\dots,L\}$

\For{$t=1,\dots,T$}
\If{$r(q_t)=1$}
\State Observe the answer $a_t$ and set $y_t\gets(q_t,a_t)$
\State Evaluate $p(y_t\mid\theta_{t-1}^i)$ for all
$i\in\{1,\dots,L\}$ according to  \eqref{logit2} and \eqref{likelihood},
\State Calculate $\{\alpha_t^i\}_{i=1}^{L}$ according to
\eqref{weight_alpha}
\State Construct $\tilde{\pi}_t(\theta\mid y_{1:t})$
according to \eqref{particle_posterior}
\State Sample $\{\theta_t^i\}_{i=1}^{L}$ from
$\tilde{\pi}_t(\theta\mid y_{1:t})$
\State Set $\alpha_t^i\gets 1/L$ for all
$i\in\{1,\dots,L\}$
\Else
\State Set
$\tilde{\pi}_t(\theta\mid y_{1:t})
\gets
\tilde{\pi}_{t-1}(\theta\mid y_{1:t-1})$
\State Set $\theta_t^i\gets\theta_{t-1}^i$ and
$\alpha_t^i\gets\alpha_{t-1}^i$ for all
$i\in\{1,\dots,L\}$
\EndIf
\EndFor

\State \Return $\tilde{\pi}_T(\theta\mid y_{1:T})$
\end{algorithmic}
\end{algorithm}

\section{Dynamic Question Designer $K$}\label{designer}

In this section, we address the problem of designing questions based on the prior distribution obtained by Algorithm 1. The objective is to select questions that provide a large amount of information about the preference parameter $\theta$, thereby enabling efficient preference estimation.

EIG is a common criterion used in experimental design and quantifies the expected amount of information obtained through a posterior update.
However, in practice, a respondent may not answer a question because of the response burden. Therefore, considering EIG alone may favor a question that is highly informative when answered but is unlikely to receive a response. To account for this issue, we incorporate the response probability into EIG and define the effective expected information gain $\mathrm{EIG}^{\mathrm{eff}}$.

\subsection{Question Design Based on Effective Expected Information Gain}\label{designer_EIG}

For a question $q_t$ at time $t$, we define the effective expected information gain $\mathrm{EIG}^{\mathrm{eff}}(q_t)$ as
\begin{align}\label{EIG_eff}
  \mathrm{EIG}^{\mathrm{eff}}(q_t)
  = \mathbb{E}[P_{r_{|q|}=1}]_t \mathrm{EIG}(q_t).
\end{align}
Thus, $\mathrm{EIG}^{\mathrm{eff}}(q_t)$ evaluates a question by jointly considering the EIG from a response and the probability of obtaining that response.

We describe the two quantities in \eqref{EIG_eff}. First, $\text{EIG}(q_t)$ is defined as
\begin{align}\label{EIG_def}
  \text{EIG}(q_t)
  &:= \mathbb{E}_{a_t \mid y_{1:t-1},q_t}
       \Bigl[
         \mathrm{KL}\bigl(
           \pi(\theta \mid y_{1:t})
           \,\big\|\,
           \pi(\theta \mid y_{1:t-1})
         \bigr)
       \Bigr].
\end{align}
EIG in \eqref{EIG_def} represents the expected information associated with the posterior update for a single question, where the expectation is taken over all possible response patterns.

Second, we model the probability of obtaining a response. Let $r_{|q|}$ be a binary variable that represents whether a response is obtained for a question consisting of $|q|$ alternatives. $r_{|q|}=1$ indicates that a response is obtained from the respondent while $r_{|q|}=0$ indicates that no response is obtained from the respondent. 
Let $P_{r_{|q|}=1}$ denote the probability that the binary variable $r_{|q|}$ takes the value 1. The probability distribution of such a binary variable is called a Bernoulli distribution. We estimate $P_{r_{|q|}=1}$ in a Bayesian manner by placing a conjugate Beta prior $\mathrm{Beta}(\alpha_{|q|}, \beta_{|q|})$ on it. The estimated $P_{r_{|q|}=1}$ at time $t$ is given as follows:
\begin{align}
  \mathbb{E}[P_{r_{|q|}=1}]_t
  =
  \frac{
    \alpha_{|q|}+N_{{|q|},1}(t)
  }{
    \alpha_{|q|}+\beta_{|q|}
    +N_{{|q|},1}(t)+N_{{|q|},0}(t)
  },
\end{align}
where $N_{|q|,1}(t)$ and $N_{|q|,0}(t)$ denote the numbers of answered and unanswered questions with $|q|$ alternatives observed up to time $t$, respectively.


Based on $\mathrm{EIG}^{\mathrm{eff}}$ in \eqref{EIG_eff}, we formulate the question design problem as follows.
\begin{problem}[Question Design Problem]\label{prob:question_design}
Let $\mathcal{Q} = \{q^{(1)}, q^{(2)}, \ldots, q^{(N)}\}$ be the candidate question set, where $q^{(i)}$ denotes the $i$-th candidate question. Find the optimal question $q_t^\ast \in \mathcal{Q}$ that maximizes $\mathrm{EIG}^{\mathrm{eff}}$:
\begin{equation}\label{eq:question_design_problem}
    q_t^\ast = \arg\max_{q^{(i)} \in \mathcal{Q}} \mathrm{EIG}^{\mathrm{eff}}(q^{(i)}).
\end{equation}
\end{problem}
Problem \ref{prob:question_design} determines a question that is likely to obtain a response and provides a large amount of information about the preference parameter $\theta$.

\begin{remark}
EIG in \eqref{EIG_def} can also be expressed using two types of entropy,
$H\bigl(a_t \mid q_t, \theta\bigr)
=-\sum_{a_t \in q_t}
\bigl[
p(a_t \mid q_t, \theta)
\log p(a_t \mid q_t, \theta)
\bigr]$
and
$H\bigl(a_t \mid y_{1:t-1},q_t\bigr)
=-\sum_{a_t \in q_t}
p(a_t \mid y_{1:t-1},q_t)
\log p(a_t \mid y_{1:t-1},q_t)$:
\begin{equation}\label{EIG_entropy}
\text{EIG}(q_t)
=
H\bigl(a_t \mid y_{1:t-1},q_t\bigr)
-
\mathbb{E}_{\theta \mid y_{1:t-1}}
\bigl[
H\bigl(a_t \mid q_t,\theta\bigr)
\bigr].
\end{equation}
As shown in \eqref{EIG_entropy}, EIG is expressed as the difference between the entropy of the response $a_t$ given the observed data
$y_{1:t-1}
=
\{(q_1,a_1),(q_2,a_2),\dots,(q_{t-1},a_{t-1})\}$
and the question $q_t$, and the expected entropy of the response $a_t$ given the preference parameter $\theta$ and the question $q_t$. Since conditional entropy represents the uncertainty of a random variable under given conditions, EIG represents how much the uncertainty about $\theta$ is reduced by the next observation $y_t$ given the data $y_{1:t-1}$ already obtained through the survey.
\end{remark}

Problem \ref{prob:question_design} belongs to the class of combinatorial optimization problems. When the total number of combinations is large, it becomes difficult to solve this problem within a finite computational time.

\subsection{Solution Algorithm}

To solve Problem \ref{prob:question_design}, a deterministic combinatorial optimization problem, we apply an $\epsilon$-greedy strategy, which is typically used in bandit problems \cite{eps_bandid}.

\subsubsection*{Overview of the Strategy}
To solve Problem \ref{prob:question_design} efficiently, we decompose Problem~\ref{prob:question_design} into two sequential subproblems:
\begin{enumerate}
    \item \textbf{Exploitation Block Identification:} The candidate question set $\mathcal{Q}$ is partitioned into $M$ blocks. By evaluating one representative question per block, this subproblem identifies a promising search region, referred to as the exploitation block. The remaining blocks are collectively defined as the exploration block.
    \item \textbf{Iterative Candidate Optimization:} Based on the exploitation block, candidate questions are sampled iteratively from the exploitation block with probability $1-\epsilon$ or the exploration block with probability $\epsilon$. This subproblem determines the approximate optimal question $\hat{q}_t^\ast$ within a given iteration budget $\tau_{\max}$.
\end{enumerate}
By sequentially solving the two subproblems, we efficiently find the approximate optimal question $\hat{q}_t^\ast$. 

\subsubsection*{Exploitation Block Identification}

We formulate the exploitation block identification. First, all candidate questions $q^{(1)}, q^{(2)}, \ldots, q^{(N)} $ are sorted in ascending order of the number of alternatives $|q|$. Then, $\mathcal{Q}$ is equally divided into $M$ blocks $\mathcal{G}_1, \ldots, \mathcal{G}_M $ satisfying $\mathcal{Q} = \bigcup_{m=1}^{M}\mathcal{G}_m$ and $\mathcal{G}_m \cap \mathcal{G}_{m'} = \emptyset$ ($m \neq m'$).
The rationale for this method of division is that questions in the same block have similar numbers of alternatives, and they are expected to yield similar $\mathrm{EIG}^{\mathrm{eff}}$ values.

Second, to evaluate the $\mathrm{EIG}^{\mathrm{eff}}$ value of each block $\mathcal{G}_m\in\{\mathcal{G}_1,\ldots,\mathcal{G}_M\}$, one question $\hat{q}_{m \mid t}$ is sampled uniformly at random from each block $\mathcal{G}_m$, and its $\mathrm{EIG}^{\mathrm{eff}}(\hat{q}_{m \mid t})$ is evaluated.

The block identification is formulated in the following problem.
\begin{subproblem}[Exploitation Block Identification]\label{prob:subproblem_A}
Find the block index $m^{\dagger} \in \{1,\ldots,M\}$ whose representative question has the highest $\mathrm{EIG}^{\mathrm{eff}}$ value:
\begin{equation}\label{eq:exploitation_block}
    m^{\dagger} = \arg\max_{m \in \{1,\ldots,M\}} \mathrm{EIG}^{\mathrm{eff}}(\hat{q}_{m \mid t}).
\end{equation}
\end{subproblem}
The block $\mathcal{G}_{m^{\dagger}}$ selected by \eqref{eq:exploitation_block} is defined as the exploitation block. The remaining blocks are collectively defined as the \emph{exploration block} $\mathcal{G}_{m^{-\dagger}} = \bigcup_{m \neq m^{\dagger}}\mathcal{G}_m$.

\subsubsection*{Iterative Candidate Optimization}
The objective of the iterative candidate optimization is to efficiently find a question with a high $\mathrm{EIG}^{\mathrm{eff}}$ by balancing exploitation and exploration. To this end, using the exploitation block $\mathcal{G}_{m^{\dagger}}$ and the exploration block $\mathcal{G}_{m^{-\dagger}}$ obtained from Problem 1-A, we perform iterative evaluations for search iterations $\tau \in \{M+1, \ldots, \tau_{\max}\}$. At each iteration $\tau$, a candidate question $\hat{q}_{\tau \mid t}$ is selected probabilistically according to the following rule:
\begin{equation}\label{eq:exploration_probability}
    P(\hat{q}_{\tau \mid t} \in \mathcal{G}_{m^{\dagger}}) = 1-\epsilon,\quad
    P(\hat{q}_{\tau \mid t} \in \mathcal{G}_{m^{-\dagger}}) = \epsilon.
\end{equation}
Within the chosen block, $\hat{q}_{\tau \mid t}$ is selected uniformly at random from the candidates that have not been evaluated.

Let $\widehat{\mathcal{Q}}_{\tau\mid t} \subseteq \mathcal{Q}$ denote the set of candidate questions retained for the final selection up to iteration $\tau$. For $\tau \in \{M+1,\ldots,\tau_{\max}\}$, $\widehat{\mathcal{Q}}_{\tau\mid t}$ consists of the question $\hat{q}_{m^{\dagger}\mid t}$ associated with the exploitation block obtained in Problem 1-A and the candidate questions evaluated from iteration $M+1$ to iteration $\tau$:
\begin{equation}
    \widehat{\mathcal{Q}}_{\tau\mid t}
    =
    \{
    \hat{q}_{m^{\dagger}\mid t},
    \hat{q}_{M+1\mid t},
    \hat{q}_{M+2\mid t},
    \ldots,
    \hat{q}_{\tau\mid t}
    \}.
\end{equation}
In particular, the set of candidate questions retained up to iteration $\tau_{\max}$ is given by
\begin{equation}
    \widehat{\mathcal{Q}}_{\tau_{\max}\mid t}
    =
    \{
    \hat{q}_{m^{\dagger}\mid t},
    \hat{q}_{M+1\mid t},
    \hat{q}_{M+2\mid t},
    \ldots,
    \hat{q}_{\tau_{\max}\mid t}
    \},
\end{equation}
which is denoted by $\widehat{\mathcal{Q}}_{t}$.

Using $\widehat{\mathcal{Q}}_t$,  Problem 1-B is defined as finding the best question among $\widehat{\mathcal{Q}}_t$:
\begin{subproblem}[Iterative Candidate Optimization]\label{prob:subproblem_B}
Let $\widehat{\mathcal{Q}}_{t}=
    \{
    \hat{q}_{m^{\dagger}\mid t},
    \hat{q}_{M+1\mid t},
    \hat{q}_{M+2\mid t},
    \ldots,
    \hat{q}_{\tau_{\max}\mid t}
    \}$. $\hat{q}_{\tau\mid t}$ is selected at random. Then, find the optimal question $\hat{q}_t^\ast$ among $\widehat{\mathcal{Q}}_t$:
\begin{equation}\label{eq:best_question}
    \hat{q}_t^\ast = \arg\max_{\hat{q}_{\tau\mid t} \in \widehat{\mathcal{Q}}_t} \mathrm{EIG}^{\mathrm{eff}}(\hat{q}_{\tau\mid t}).
\end{equation}
\end{subproblem}

\subsubsection*{Algorithm Summary}
The overall solution method combining Problem 1-A and Problem 1-B is summarized in Algorithm~\ref{alg:epsilon_greedy_question_optimization}.

\begin{algorithm}[t]
\caption{Question Design by $\epsilon$-Greedy Strategy}
\label{alg:epsilon_greedy_question_optimization}
\begin{algorithmic}[1]
\Require Candidate question set $\mathcal{Q}$, number of blocks $M$, maximum search iterations $\tau_{\max}$, and exploration probability $\epsilon$
\Ensure Approximate optimal question $\hat{q}_t^\ast$ at estimation time $t$

\State Partition $\mathcal{Q}$ into $\mathcal{G}_1,\ldots,\mathcal{G}_M$

\For{$m=1,\ldots,M$}
    \State Select one question $\hat{q}_{m\mid t} \in \mathcal{G}_m$ uniformly at random
    \State Evaluate $\mathrm{EIG}^{\mathrm{eff}}(\hat{q}_{m\mid t})$
\EndFor

\State Determine exploitation block $\mathcal{G}_{m^{\dagger}}$ according to \eqref{eq:exploitation_block}
\State Construct exploration block
$\mathcal{G}_{m^{-\dagger}}
=
\bigcup_{m\neq m^{\dagger}}\mathcal{G}_m$

\State Initialize
$\widehat{\mathcal{Q}}_{M\mid t}
\gets
\{\hat{q}_{m^{\dagger}\mid t}\}$

\For{$\tau=M+1,\ldots,\tau_{\max}$}
    \State Draw $u_\tau \sim \mathrm{Uniform}(0,1)$

    \If{$u_\tau < \epsilon$}
        \State Select
        $\hat{q}_{\tau\mid t}
        \in
        \mathcal{G}_{m^{-\dagger}}
        \setminus
        \{\hat{q}_{1\mid t},\ldots,\hat{q}_{\tau-1\mid t}\}$
        uniformly at random
    \Else
        \State Select
        $\hat{q}_{\tau\mid t}
        \in
        \mathcal{G}_{m^{\dagger}}
        \setminus
        \{\hat{q}_{1\mid t},\ldots,\hat{q}_{\tau-1\mid t}\}$
        uniformly at random
    \EndIf

    \State Evaluate
    $\mathrm{EIG}^{\mathrm{eff}}(\hat{q}_{\tau\mid t})$

    \State Update
    $\widehat{\mathcal{Q}}_{\tau\mid t}
    \gets
    \widehat{\mathcal{Q}}_{\tau-1\mid t}
    \cup
    \{\hat{q}_{\tau\mid t}\}$
\EndFor

\State Set
$\widehat{\mathcal{Q}}_t
\gets
\widehat{\mathcal{Q}}_{\tau_{\max}\mid t}$

\State Determine
$\hat{q}_t^\ast$
from $\widehat{\mathcal{Q}}_t$
according to \eqref{eq:best_question}

\State \Return $\hat{q}_t^\ast$
\end{algorithmic}
\end{algorithm}

\section{Theoretical Analysis}\label{analysis}
In this section, we theoretically analyze the estimation accuracy of the respondent preference estimation system designed up to Sections~\ref{esitimator} and \ref{designer}. As an accuracy metric, we use Bayes risk and estimate how small Bayes risk can become through the method in Algorithm 1. Bayes risk is widely used as a metric for evaluating the error of Bayesian estimation. We derive a lower bound on Bayes risk using the multivariate van Trees inequality\cite{bound}.
In addition, in this section,  we derive the probability that the question design through the $\epsilon$-greedy strategy considered in Section 4 obtains the true optimal solution.

\subsection{Derivation of a Lower Bound on Bayes Risk}
In this subsection, we derive a lower bound on the Bayes risk of the estimated preference $\hat{\theta}$, which is the target of estimation. For the estimated preference $\hat{\theta}$, the following Theorem 1 holds.

\begin{theorem}
Assume that $\theta_i\ge\varepsilon$ holds $\forall$ $i$. Then, letting $\Theta=\{ \theta \in \mathbb{R}^{n} \mid \sum_{i=1}^{n} \theta_i =1 \}$, the following inequality holds.
\begin{equation}
\int_{\Theta}\mathbb{E}_{y_{1:T}\mid\theta}[| \hat{\theta} - \theta |^2 ]\pi_0(\theta) d\theta \ge 
\frac{(n-1)^2}{\dfrac{4(n-1)T}{\varepsilon^2}+\tilde I(\pi_0)},
\label{eq:bayesrisk_lower}
\end{equation}
where $\tilde I(\pi_0)$ is the trace of the Fisher information matrix $I(\pi_0)$ of the prior distribution $\pi_0$.
\end{theorem}

To prove Theorem 1, we prepare the following two lemmas. Lemma 1 follows directly from the multivariate van Trees inequality.

Before presenting Lemma 1, we define a parameter $\phi$ whose components are independent as follows:
\begin{equation}\label{phi_theta}
\phi = (\phi_1, \dots, \phi_{n-1})^T \in \mathbb{R}^{n-1} , \phi_i = \theta_i.
\end{equation}
The domain of $\phi$, denoted by $\Phi$, is defined as follows:
\begin{equation}
\Phi = \{ \phi \in \mathbb{R}^{n-1} \mid \sum_{i=1}^{n-1} \phi_i < 1 \}.
\end{equation}

Moreover, assume that the likelihood function $p(y_{1:T}|\phi)$ is measurable with respect to $\phi$ and $y$, and that the prior distribution $\pi_0(\phi)$ is a smooth probability density function on the domain $\Phi$ of class $C^1$. On the boundary of $\Phi$, $\pi_0(\phi)=0$. Furthermore, assume that, for the observations $y_{1:T}$, the Fisher information matrix $I_{1:T}(\phi)$ is able to be defined, and that $\text{diag}\{{I_{1:T}(\phi)}\}^{1/2}$ is locally integrable.
Under the above assumptions, the following Lemma 1 holds.
\begin{lemma}
For the domain $\Phi$ and $\phi \in \Phi$, the following inequality holds \cite {bound}.
\begin{equation}
    \label{eq:vanTrees}
    \int_{\Phi}\mathbb{E}_{y_{1:T}\mid\phi}[| \hat{\phi} - \phi |^2] \pi_0(\phi) d\phi 
    \ge 
    \frac{ \left( \int_{\Phi} \mathrm{div } \phi \cdot \pi_0(\phi) d\phi \right)^2 }
    { \int_{\Phi} \mathrm{Tr}\left( I_{1:T}(\phi) \right) \pi_0(\phi) d\phi + \tilde{I}(\pi_0) },
\end{equation}
where $ I_{1:T}$ is defined by \eqref{total_fisher_information}
\begin{align}
I_{1:T}(\phi)
:=
\mathbb{E}_{y_{1:T}\mid\phi}
\left[
\nabla_{\phi}\log p(y_{1:T}\mid\phi)
\nabla_{\phi}\log p(y_{1:T}\mid\phi)^{T}
\right]
\label{total_fisher_information}
\end{align}
\end{lemma}

\begin{lemma}
Assume that, for any $a_t\in q_t$, $p(a_t\mid q_t,\phi)>0$
and that $p(a_t\mid q_t,\phi)$ is differentiable with respect to $\phi$. Then, the Fisher information matrix for all observations $y_{1:T}$ is computed as follows:
\begin{equation}
I_{1:T}(\phi)
=
\sum_{t=1}^{T}
\mathbb{E}_{y_{1:t-1}\mid\phi}
\left[
I_t(\phi\mid y_{1:t-1})
\right],
\label{eq:fisher_information_additivity}
\end{equation}
where $ I_{1:T}$ is defined by \eqref{total_fisher_information}
\end{lemma}
The proof of Lemma 2 is given in Appendix.

\subsubsection*{Proof of Theorem $1$}
We explicitly evaluate the numerator and denominator on the right-hand side of \eqref{eq:vanTrees} in Lemma 1.

First, we compute the numerator. $\text{div } \phi$ in the numerator of \eqref{eq:vanTrees} is computed as follows:
\begin{equation}
    \text{div } \phi = \sum_{i=1}^{n-1} \frac{\partial \phi_i}{\partial \phi_i} = \sum_{i=1}^{n-1} 1 = n - 1.
\end{equation}
Therefore, since the integral of the prior distribution over the entire space is equal to 1, the entire numerator is reduced to
\begin{equation}\label{Numerator}
    \left( (n-1) \int_{\Phi} \pi_0(\phi) d\phi \right)^2 = (n-1)^2.
\end{equation}

Next, we evaluate the denominator on the right-hand side of \eqref{eq:vanTrees}. First, applying Lemma 2, the first term in the denominator is expanded as follows:
\begin{align}
    &\int_{\Phi} \text{Tr}\left( I_{1:T}(\phi) \right) \pi_0(\phi) d\phi\nonumber\\ 
    &= \int_{\Phi} \left(\text{Tr}\left(\sum_{t=1}^{T}
\mathbb{E}_{y_{1:t-1}\mid\phi}
\left[
I_t(\phi\mid y_{1:t-1})
\right] \right)\right) \pi_0(\phi) d\phi\nonumber\\
&=\int_{\Phi} \left(\sum_{t=1}^{T} \mathbb{E}_{y_{1:t-1}\mid\phi} \left[ \text{Tr}\left(I_t(\phi\mid y_{1:t-1}) \right) \right] \right) \pi_0(\phi) d\phi.
\label{Information_1:T}
\end{align}
We evaluate $\text{Tr}\left(I_t(\phi\mid y_{1:t-1}) \right)$ in \eqref{Information_1:T}.
The conditional probability of $y_t=(q_t,a_t)$, $p(y_t\mid y_{1:t-1},\phi)$, is decomposed as follows:
\begin{align}
&p(y_t\mid y_{1:t-1},\phi)\nonumber\\
&=
p(q_t,a_t\mid y_{1:t-1},\phi)
\nonumber\\
&=
p(q_t\mid y_{1:t-1},\phi)
p(a_t\mid q_t,y_{1:t-1},\phi)
\nonumber\\
&=
p(q_t\mid y_{1:t-1})
p(a_t\mid q_t,\phi).
\label{conditional_observation_factorization}
\end{align}
Taking the logarithm of both sides of \eqref{conditional_observation_factorization} and differentiating with respect to $\phi$, since $p(q_t\mid y_{1:t-1})$ does not depend on $\phi$, we obtain
\begin{align}
\nabla_{\phi}
&\log p(y_t\mid y_{1:t-1},\phi)\nonumber\\
&=
\nabla_{\phi}
\log p(q_t\mid y_{1:t-1})
+
\nabla_{\phi}
\log p(a_t\mid q_t,\phi)
\nonumber\\
&=
\nabla_{\phi}
\log p(a_t\mid q_t,\phi).
\label{conditional_gradient_reduction}
\end{align}
Let
\begin{equation}
    L_t(\phi) := \log p(a_t\mid q_t,\phi)=\log \theta_{a_t}(\phi) - \log ( \sum_{j \in q_t} \theta_j(\phi)  ).
\end{equation}
Then, the Fisher information matrix $I_t(\phi\mid y_{1:t-1})$ in \eqref{total_fisher_information} is written as follows:
\begin{equation}
\begin{aligned}
    &I_t(\phi\mid y_{1:t-1})\nonumber\\ 
    &= \mathbb{E}_{y_t\mid y_{1:t-1},\phi}
\left[
\nabla_{\phi}L_t(\phi)
\nabla_{\phi}L_t(\phi)^{T}\right]\\
&= \mathbb{E}_{y_t\mid y_{1:t-1},\phi}
    \begin{pmatrix}
        \left(\frac{\partial L_t}{\partial \phi_1}\right)^2 & \frac{\partial L_t}{\partial \phi_1}\frac{\partial L_t}{\partial \phi_2} & \dots & \frac{\partial L_t}{\partial \phi_1}\frac{\partial L_t}{\partial \phi_{n-1}} \\
        \frac{\partial L_t}{\partial \phi_2}\frac{\partial L_t}{\partial \phi_1} & \left(\frac{\partial L_t}{\partial \phi_2}\right)^2 & \dots & \frac{\partial L_t}{\partial \phi_2}\frac{\partial L_t}{\partial \phi_{n-1}} \\
        \vdots & \vdots & \ddots & \vdots \\
        \frac{\partial L_t}{\partial \phi_{n-1}}\frac{\partial L_t}{\partial \phi_1} & \frac{\partial L_t}{\partial \phi_{n-1}}\frac{\partial L_t}{\partial \phi_2} & \dots & \left(\frac{\partial L_t}{\partial \phi_{n-1}}\right)^2
    \end{pmatrix}.
\end{aligned}
\end{equation}
Furthermore, for $I_t(\phi\mid y_{1:t-1})$, its trace $\text{Tr}(I_t(\phi\mid y_{1:t-1}))$ is written as follows:
\begin{equation}\label{TraceI}
    \text{Tr}(I_t(\phi\mid y_{1:t-1})) = \mathbb{E}_{y_t\mid y_{1:t-1},\phi}\left[\sum_{i=1}^{n-1} \left( \frac{\partial L_t}{\partial \phi_i} \right)^2\right].
\end{equation}
Applying the chain rule, the derivative of $L_t$ with respect to $\phi_i$ is transformed into a derivative with respect to $\theta$. Using $\frac{\partial \theta_i}{\partial \phi_i} = 1$, $\frac{\partial \theta_j}{\partial \phi_i} = 0$, and $\frac{\partial \theta_n}{\partial \phi_i} = -1$, we obtain
\begin{align}
    \frac{\partial L_t}{\partial \phi_i} 
&=\sum_{j=1}^{n}\frac{\partial L_t}{\partial \theta_j} \frac{\partial \theta_j}{\partial \phi_i}\nonumber\\
&= \frac{\partial L_t}{\partial \theta_i} \frac{\partial \theta_i}{\partial \phi_i} + \frac{\partial L_t}{\partial \theta_n} \frac{\partial \theta_n}{\partial \phi_i} \nonumber\\
    &= \frac{\partial L_t}{\partial \theta_i} - \frac{\partial L_t}{\partial \theta_n}.
\end{align}
Noting that $(a-b)^2 \le 2(a^2 + b^2)$ holds for all real values a and b, we obtain 
\begin{align}\label{extention_Lphi}
\sum_{i=1}^{n-1} \left( \frac{\partial L_t}{\partial \phi_i} \right)^2
&=\sum_{i=1}^{n-1} \left[ \left( \frac{\partial L_t}{\partial \theta_i} - \frac{\partial L_t}{\partial \theta_n} \right)^2 \right]\nonumber\\
&\le \sum_{i=1}^{n-1} 2 \left[ \left( \frac{\partial L_t}{\partial \theta_i} \right)^2 + \left( \frac{\partial L_t}{\partial \theta_n} \right)^2 \right] \nonumber\\
    &= 2 \sum_{i=1}^{n-1}  \left[ \left( \frac{\partial L_t}{\partial \theta_i} \right)^2 \right]+2(n-1)\left( \frac{\partial L_t}{\partial \theta_n} \right)^2.
\end{align}
$\frac{\partial L_t}{\partial \theta_i}$ is computed as follows:
\begin{equation}\label{eq:dl_dtheta_i}
  \frac{\partial L_t}{\partial \theta_i}
  =
  \begin{cases}
    \dfrac{1}{\theta_{a_t}}
    - \dfrac{1}{\sum_{j\in q_t}\theta_j},
      & (i=a_t), \\[10pt]
    - \dfrac{1}{\sum_{j\in q_t}\theta_j},
      & (i\in q_t,\ i\ne a_t), \\[10pt]
    0, & (i\notin q_t).
  \end{cases}
\end{equation}
Under the assumption $\theta_{i}\ge\varepsilon$, we obtain
\begin{equation}\label{eq:dl_bound}
  \left|\frac{\partial L_t}{\partial \theta_i}\right|
  \le \frac{1}{\varepsilon}.
\end{equation}
By applying \eqref{extention_Lphi} and \eqref{eq:dl_bound} to \eqref{TraceI}, we obtain
\begin{align}\label{TrI_t}
&\text{Tr}(I_t(\phi\mid y_{1:t-1}))\nonumber\\&=\mathbb{E}_{y_t\mid y_{1:t-1},\phi}\left[\sum_{i=1}^{n-1} \left( \frac{\partial L_t}{\partial \phi_i} \right)^2\right]\nonumber\\
&\le\mathbb{E}_{y_t\mid y_{1:t-1},\phi}\left[2 \sum_{i=1}^{n-1}  \left[ \left( \frac{\partial L_t}{\partial \theta_i} \right)^2 \right]+2(n-1)\left( \frac{\partial L_t}{\partial \theta_n} \right)^2\right]\nonumber\\
&\le\frac{4(n-1)}{\varepsilon^2}.
\end{align}
From \eqref{Information_1:T} and \eqref{TrI_t}, the following inequality holds.
\begin{align}\label{int_Tr_I}
    &\int_{\Phi} \text{Tr}\left( I_{1:T}(\phi) \right) \pi_0(\phi) d\phi\nonumber\\
    &=\int_{\Phi} \left(\sum_{t=1}^{T} \mathbb{E}_{y_{1:t-1}\mid\phi} \left[ \text{Tr}\left(I_t(\phi\mid y_{1:t-1}) \right) \right] \right) \pi_0(\phi) d\phi\nonumber\\
    & \le \int_{\Phi} T\left( \frac{4(n-1)}{\varepsilon^2} \right) \pi_0(\phi) d\phi\nonumber\\
    &=\left( \frac{4(n-1)T}{\varepsilon^2} \right)\int_{\Phi} \pi_0(\phi) d\phi\nonumber\\ 
    &=\frac{4(n-1)T}{\varepsilon^2}.
\end{align}

From \eqref{eq:vanTrees}, \eqref{Numerator}, and \eqref{int_Tr_I}, the following inequality holds.
\begin{align}\label{van_exp}
    \int_{\Phi}\mathbb{E}_{y_{1:T}\mid\phi}[| \hat{\phi} - \phi |^2] \pi_0(\phi) d\phi 
    &\ge\frac{ \left( n-1\right)^2 }
    { \frac{4(n-1)T}{\varepsilon^2} + \tilde{I}(\pi_0) }.
\end{align}

Furthermore, for the preference estimation error $|\hat{\theta} - \theta|^2$, since \eqref{phi_theta} implies that $|\hat{\theta} - \theta|^2=|\hat{\phi} - \phi |^2+(\hat{\theta}_n - \theta_n)^2\ge|\hat{\phi} - \phi |^2$ holds, the following inequality immediately holds.
\begin{align}\label{theta_phi}
    \int_{\Theta}\mathbb{E}_{y_{1:T}\mid\theta}[| \hat{\theta} - \theta |^2] \pi_0(\theta) d\theta  \ge \int_{\Phi}\mathbb{E}_{y_{1:T}\mid\phi}| \hat{\phi} - \phi |^2 \pi_0(\phi) d\phi,\\
    \Theta=\{ \theta \in \mathbb{R}^{n} \mid \sum_{i=1}^{n} \theta_i =1 \}.\nonumber
\end{align}

From \eqref{van_exp} and \eqref{theta_phi}, \eqref{eq:bayesrisk_lower} holds.
This completes the proof of the theorem.

\subsection{Probability of Obtaining the True Optimal Solution by $\epsilon$-Greedy Strategy}
We derive the probability that the true optimal solution is obtained by the $\epsilon$-greedy strategy within $\tau$ iterations, denoted by $P_{\mathrm{success}}$. In the derivation, recall that the meanings of the symbols are summarized as follows: $N$ is the number of candidate questions to be searched, $M$ is the number of blocks into which the search space $\mathcal{Q}$ is divided, $\tau$ is the number of optimization iterations, $q^{\ast}$ is the true optimal solution, $\mathcal{G}^{\ast}$ is the block containing the optimal solution, and $\epsilon$ is the exploration probability in the $\epsilon$-greedy strategy.

First, we prepare the following assumption, which is necessary for deriving $P_{\mathrm{success}}$.
We denote the Gumbel distribution with location parameter $\mu$ and scale parameter $\beta$ by $\mathrm{Gumbel}(\mu,\beta)$. Its probability density function is given by
\begin{equation}
\label{eq:gumbel_distribution}
f(x;\mu,\beta)
=
\frac{1}{\beta}
\exp\left(-\frac{x-\mu}{\beta}\right)
\exp\left(
-\exp\left(-\frac{x-\mu}{\beta}\right)
\right).
\end{equation}

\begin{assumption}
Let $X_1$ denote the $\mathrm{EIG}^{\mathrm{eff}}$ value in $\mathcal{G}^{\ast}$, and let $X_i$ denote the $\mathrm{EIG}^{\mathrm{eff}}$ values in the other blocks. We assume that
\begin{equation}
\label{eq:mtg04}
X_1 \sim \mathrm{Gumbel}(\mu_1,\beta)
\end{equation}
and
\begin{equation}
\label{eq:mtg05}
X_i \sim \mathrm{Gumbel}(\mu_0,\beta),
\end{equation}
where $\mu_1$, $\mu_0$, and $\beta$ are positive constants. The pdf of $X_1$ is denoted by $f_1$ and The pdf of $X_i$ is denoted by $f_i$.

\end{assumption}

Second, we prepare the following Lemma 3. 
\begin{lemma}
Assumption $2$ holds. Then, the probability $\gamma$ that the $\mathrm{EIG}^{\mathrm{eff}}$ value of $\mathcal{G}^{\ast}$, $X_1$, is greater than or equal to all the $\mathrm{EIG}^{\mathrm{eff}}$ values of the other $M-1$ blocks, $X_2,..., X_i,..., X_M$, is described as follows:
\begin{equation}
\label{eq:mtg03'}
\gamma=\frac{e^d}{e^d+M-1},
\end{equation}
where $d:=\frac{\mu_1-\mu_0}{\beta}$.
\end{lemma}
The proof of Lemma 3 is given in Appendix.

Under Assumption 2 and Lemma 3, the following theorem holds.
\begin{theorem}
Under Assumption 2 and Lemma 3, the probability that the true optimal solution is obtained by the $\epsilon$-greedy strategy within $\tau \in [M, \frac{N}{M} + M - 1]$ iterations, denoted by $P_{\mathrm{success}}$, is described as follows:
\begin{align}\label{eq:mtg01}
&P_{\mathrm{success}}\nonumber\\
&=\left(1-\frac{M}{N}\right)
\left(
\frac{M(1-\epsilon)\gamma}{N-M}
+
\frac{M(1-\gamma)\epsilon}{NM-N-M(M-1)}
\right)\tau \nonumber\\
&\quad +\frac{M}{N}\nonumber\\
&\quad-
\left(1-\frac{M}{N}\right)
\left(
\frac{M^2(1-\epsilon)\gamma}{N-M}
+
\frac{M^2(1-\gamma)\epsilon}{NM-N-M(M-1)}
\right).
\end{align}
\end{theorem}

\subsubsection*{Proof of Theorem $2$}
First, for $0\le\tau\le M$, when one candidate is searched from each group, the probability that $q^{\ast}$ is not obtained, denoted by $P_{\text{fir}}$, is as follows:
\begin{equation}\label{eq:mtg22}
P_{\text{fir}}=1-\frac{M}{N}.
\end{equation}

Next, consider the probability that, after missing $q^{\ast}$ in the first $M$ searches, $q^{\ast}$ is still missed in the subsequent $\tau-M$ searches, denoted by $P_{\text{sec}}$. $P_{\text{sec}}$ must be considered by dividing it into two cases: (1) the case where $\mathcal{G}^{\ast}$ is selected as the exploitation block and $q^{\ast}$ is missed, and (2) the case where $\mathcal{G}^{\ast}$ is not selected as the exploitation block and $q^{\ast}$ is missed. $P_{\text{sec}}$ is described as the sum of the probability $P^1_{\text{sec}}$ in case (1) and the probability $P^2_{\text{sec}}$ in case (2), as follows:
\begin{equation}
\label{eq:mtg23}
P_{\text{sec}}=P^1_{\text{sec}}+P^2_{\text{sec}}.
\end{equation}

\subsubsection*{(1) The case where $\mathcal{G}^{\ast}$ is selected as the exploitation block, exploited $x$ times, explored $\tau-M-x$ times, and $q^{\ast}$ is missed}

In case (1), since $q^{\ast}$ was missed in the initial stage, $q^{\ast}$ is one of the $K-1$ unevaluated candidates in $\mathcal{G}^*$; and since selection within a block is without replacement, the probability that $x$ exploitation draws miss $q^{\ast}$ is exactly $1 - \frac{x}{K-1}$.\footnote{Please note that $x \le \tau - M \le K-1$.} Exploration draws cannot find $q^{\ast}$ because $q^{\ast} \in G^* = G_{m^\dagger}$. By using \eqref{eq:mtg03'} in Lemma 3, $P^1_{\text{sec}}$ that $q^{\ast}$ is missed after exploiting $x$ times and exploring $\tau-M-x$ times is expressed as follows: 
\begin{align}\label{eq:mtg24}
P^1_{\text{sec}}=\gamma&\sum_{x=0}^{\tau-M}{}_{\tau-M}C_x(1-\epsilon)^x\epsilon^{\tau-M-x}
\left(1-\frac{x}{K-1}\right) \nonumber\\
=\gamma&\sum_{x=0}^{\tau-M}{}_{\tau-M}C_x(1-\epsilon)^x\epsilon^{\tau-M-x}\nonumber\\
&-\gamma\sum_{x=0}^{\tau-M}{}_{\tau-M}C_x(1-\epsilon)^x\epsilon^{\tau-M-x}\frac{x}{K-1} \nonumber\\
=\gamma&-\frac{\gamma}{K-1}(\tau-M)(1-\epsilon),
\end{align}
where $K=N/M$.

\subsubsection*{(2) The case where $\mathcal{G}^{\ast}$ is not selected as the exploitation block}
In case (2), $q^{\ast}$ lies in the exploration block, which contains $N - K$ candidates in total, of which $M-1$ have already been evaluated in the initial stage; hence $q^{\ast}$ is one of $N - K - (M-1)$ unevaluated exploration candidates, and only exploration draws can find it. By the same without-replacement argument as in case (1), the probability that $q^{\ast}$ is missed after exploring $x$ times and exploiting $\tau-M-x$ times is expressed as follows:
\begin{align}\label{eq:mtg26}
P^2_{\text{sec}}=(1-\gamma)
-\frac{1-\gamma}{N-K-(M-1)}(\tau-M)\epsilon.
\end{align}

Therefore, substituting \eqref{eq:mtg24} and \eqref{eq:mtg26} into \eqref{eq:mtg23}, the probability $P_{\text{sec}}$ that, after missing $q^{\ast}$ in the first $M$ searches, $q^{\ast}$ is still missed in the subsequent $\tau-M$ searches is described as follows.
\begin{align}\label{eq:mtg28}
P_{\mathrm{sec}}
&=\gamma+(1-\gamma)
-\frac{\gamma}{K-1}(\tau-M)(1-\epsilon)\nonumber\\
&\quad-\frac{1-\gamma}{N-K-(M-1)}(\tau-M)\epsilon \nonumber\\
&=1-\frac{\gamma}{K-1}(\tau-M)(1-\epsilon)\nonumber\\
&\quad-\frac{1-\gamma}{N-K-(M-1)}(\tau-M)\epsilon.
\end{align}

$P_{\mathrm{success}}$ is expressed as follows:
\begin{equation}
\label{eq:mtg29}
P_{\mathrm{success}}=1-P_{\mathrm{fir}}P_{\mathrm{sec}}.
\end{equation}

Therefore, substituting \eqref{eq:mtg22} and \eqref{eq:mtg28} into \eqref{eq:mtg29} and using $K=N/M$, we finally obtain the following equation:
\begin{align}\label{eq:mtg30}
&P_{\mathrm{success}}\nonumber\\
&=\left(1-\frac{M}{N}\right)
\left(
\frac{M(1-\epsilon)\gamma}{N-M}
+
\frac{M(1-\gamma)\epsilon}{NM-N-M(M-1)}
\right)\tau \nonumber\\
&\quad +\frac{M}{N}\nonumber\\
&\quad-
\left(1-\frac{M}{N}\right)
\left(
\frac{M^2(1-\epsilon)\gamma}{N-M}
+\frac{M^2(1-\gamma)\epsilon}{NM-N-M(M-1)}\right).
\end{align}
Theorem 2 follows from \eqref{eq:mtg30}.

\begin{remark}
Many analyses of an $\epsilon$-greedy strategy have been evaluated using inequalities, such as upper and lower bounds. In contrast, few studies have derived the probability as an equality. Therefore, the derivation in this paper is able to be regarded as a new attempt.
The key idea in the derivation of Theorem $2$ is to assume that the objective function values follow a Gumbel distribution. In theoretical analyses of bandit algorithms, such as the $\epsilon$-greedy strategy, a Gaussian distribution is usually assumed for the values \cite{bandid_gauss}. However, a Gaussian distribution is not convenient for analytical derivation because its CDF cannot be written in a simple closed form.
For this reason, many theoretical studies based on a Gaussian distributions derive bounds, such as lower bounds on the regret with respect to the true optimal value. In contrast, this paper assumes a Gumbel distribution. This assumption enables us to derive the probability of obtaining the true optimal solution as an equality.
\end{remark}

Furthermore, the validity of Theorem 2 was confirmed through numerical experiments. First, we construct a combinatorial optimization problem whose objective function values are randomly generated from Gumbel distributions. The values in $\mathcal{G}^{\ast}$, which contains the optimal solution, are generated from a Gumbel distribution with location parameter 2.0 and scale parameter 1.0. The values outside $\mathcal{G}^{\ast}$ are generated from a Gumbel distribution with location parameter 0 and scale parameter 1.0. We solve this combinatorial optimization problem by the $\epsilon$-greedy strategy with exploration rate $\epsilon=0.2$. Then, we visualize the transition of the success rate over the search iterations. We also check whether the simulation results agree with the straight line given by \eqref{eq:mtg01} within the iteration range specified by Theorem 2. The number of simulations is set to 100, 1000, 5000, and 10000. For each case, the transition of the success rate over the search iterations is visualized.

\begin{figure*}[t]
\centering
\includegraphics[width=1\linewidth]{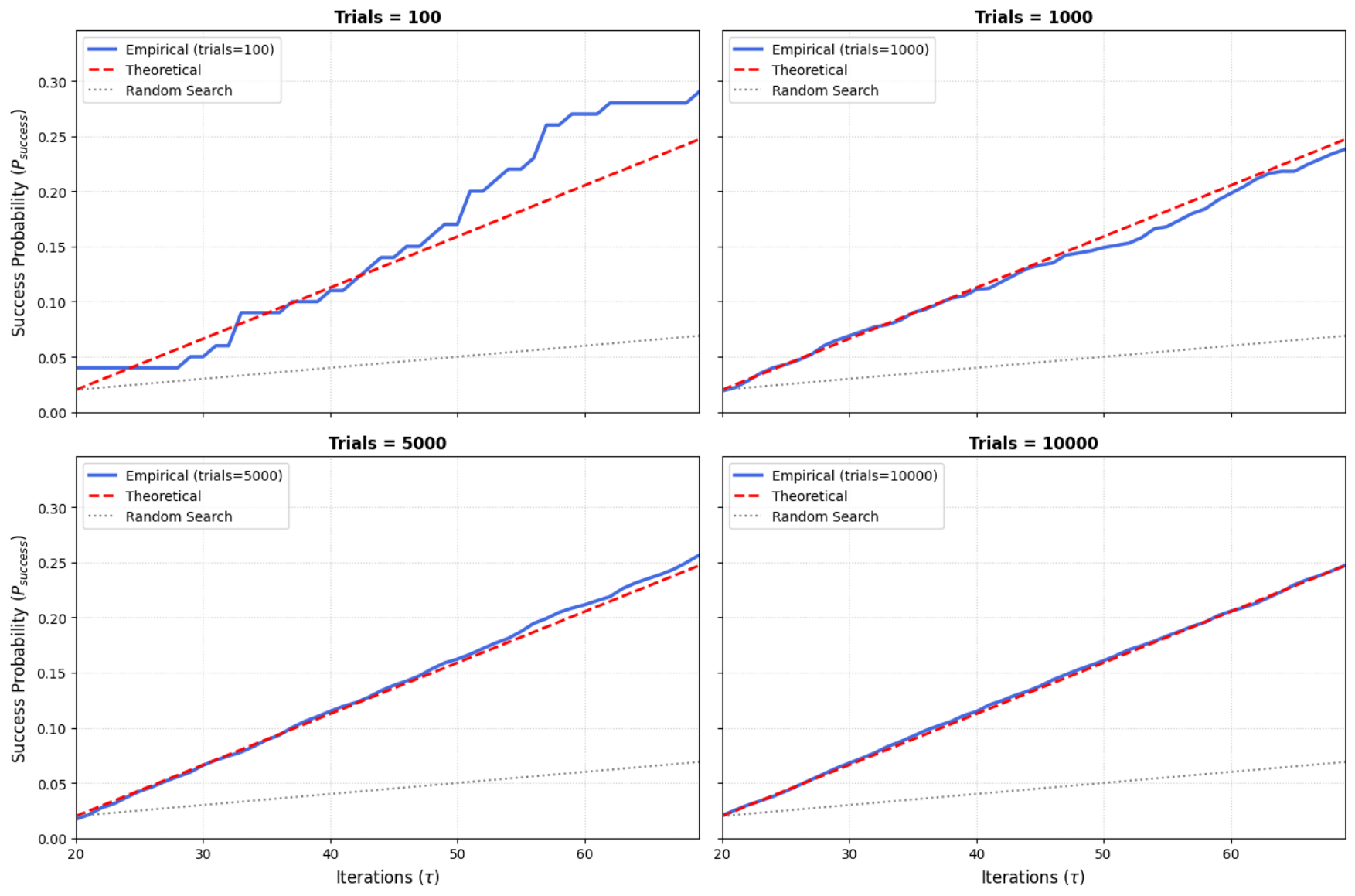}
\caption{Transition of the correct-answer rate over the search iterations and theoretical value of the correct-answer probability for each number of simulations}
\label{P_success}
\end{figure*}

The results are shown in Fig.~\ref{P_success}.
In Fig.~\ref{P_success}, the blue line represents the simulation result. The red dotted line represents the theoretical value of $P_{\mathrm{success}}$ given by \eqref{eq:mtg01}. The black dotted line represents the theoretical value for random search.

From Fig.~\ref{P_success}, we can see that the blue line approaches the red dotted line as the number of simulations increases. This means that the theoretical value given by \eqref{eq:mtg01} accurately represents the simulation result under Assumption 2. Therefore, the validity of Theorem 2 was verified not only theoretically but also experimentally. In addition, the blue line is above the black dotted line. This shows that the $\epsilon$-greedy strategy achieves better performance than random search for the same number of iterations.

\section{Simulation Experiment}

In this section, we conduct a simulation experiment to evaluate the performance of the estimator $E$ and the question designer $K$.

In the experiment, we compare four question design methods.
The first method selects questions randomly.
The second method selects questions by maximizing EIG.
The third method selects questions by maximizing $\mathrm{EIG}^{\mathrm{eff}}$ by exhaustive search.
The fourth method selects questions by maximizing $\mathrm{EIG}^{\mathrm{eff}}$ by the $\epsilon$-greedy strategy.
The number of alternatives $n$ is set to $7$, the final estimation time $T$ is set to $100$, and the number of particles $L$ is set to $2000$.
The number of iterations in the $\epsilon$-greedy strategy $\tau_{max}$ is set to $80$.
This corresponds to approximately $66\%$ of the total number of candidate questions, $N=120$. For these four question design methods, we examine the time evolution of the estimation error.

We also visualize the lower bound of the Bayes risk derived in Section~\ref{analysis}. By confirming that this lower bound remains below the estimation error, we experimentally verify the validity of Theorem 1.

\begin{figure}[t]
\centering
\includegraphics[width=1\linewidth]{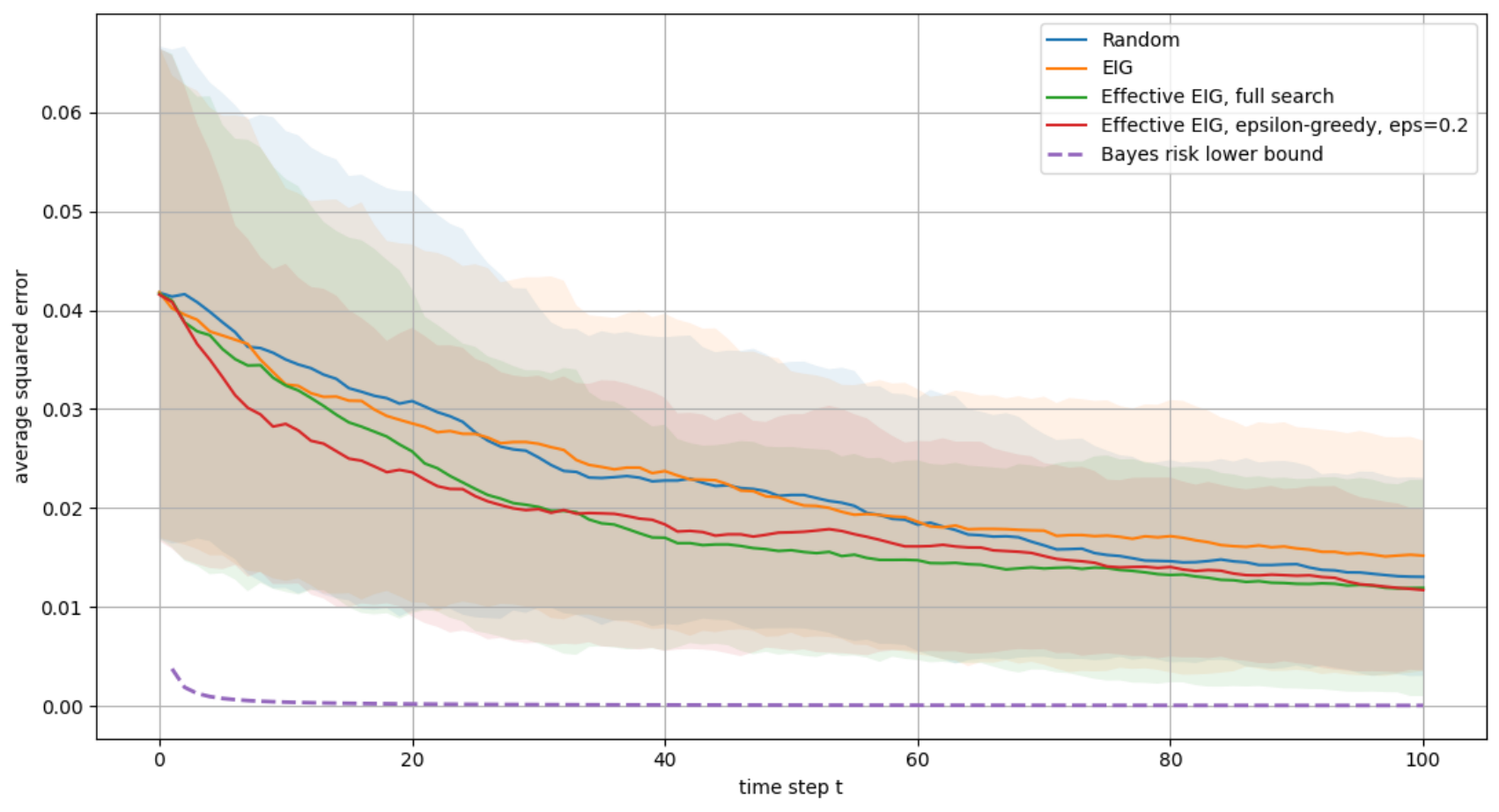}
\caption{Comparison of squared estimation errors for each method and the lower bound of the Bayes risk}
\label{fig:simu}
\end{figure}

Figure \ref{fig:simu} shows the time evolution of the squared estimation error between the estimated preference $\hat{\theta}$ and the true preference $\theta$. The figure enables a comparison of the four methods.
In addition, the final squared estimation errors for each method in Fig.~\ref{fig:simu} are summarized in Table~1.
\begin{table}[t]
  \centering
  \caption{Comparison of final squared estimation errors for each method}
  \label{tab:final_average_squared_error}
  \begin{tabular}{lc}
    \hline
    Method & Final squared estimation error \\
    \hline
    Random & 0.01304 \\
    EIG & 0.01519 \\
    Effective EIG, full search & 0.01192 \\
    Effective EIG, $\epsilon$-greedy & 0.01171 \\
    \hline
  \end{tabular}
\end{table}
The results show that the proposed method, which designs questions by maximizing $\mathrm{EIG}^{\mathrm{eff}}$, achieves the greatest reduction in squared estimation error.
In addition, the actual computation time from $t=0$ to $T$ was 2.76 seconds when the maximization problem of $\mathrm{EIG}^{\mathrm{eff}}$ was solved by full search, whereas it was 2.28 seconds when it was solved by the $\epsilon$-greedy strategy. Moreover, comparing the squared estimation errors of these two cases in Table~1 shows that there is no significant difference between the full-search method and $\epsilon$-greedy strategy. This result confirms that the estimation accuracy does not significantly decrease even when question optimization is performed approximately by the $\epsilon$-greedy strategy.

Furthermore, the lower bound of the Bayes risk derived in Section~\ref{analysis} is below the estimation errors of all four question design methods. This result supports the validity of the theoretical result.

\section{Conclusion}
This paper proposed a method for estimating respondents' preferences from questionnaire responses and a method for optimally designing questionnaire questions. First, we proposed a preference model for respondents. Next, based on the model, we proposed a method for estimating the true preferences using Bayesian estimation. Furthermore, to reduce the computational cost of Bayesian estimation, we used a particle filter. We then formulated the question design problem as an $\mathrm{EIG}^{\mathrm{eff}}$ maximization problem. In addition, by using the $\epsilon$-greedy strategy to solve the question design problem, we reduced the computational cost required for question design. Finally, we investigated the performance of both the estimation method and the question design method through theoretical analysis and simulations. As a result, it was confirmed that question design using the $\epsilon$-greedy strategy achieves performance comparable to that of full search. In the theoretical analysis, by assuming a Gumbel distribution as the distribution followed by the $\mathrm{EIG}^{\mathrm{eff}}$, we analytically derived the probability of obtaining the true optimal solution. Moreover, regarding the analysis of estimation error, we found that a lower bound on the Bayes risk is able to be analytically derived even when the observations $y_{1:T}$ are non-independent and obtained from different distributions.




\appendix
\section{Appendix: Proof of Lemma 2}
First, as a preliminary, recall that the question $q_t$ is determined based on the past observation history $y_{1:t-1}$. Then, $q_t$ does not directly depend on $\phi$. That is,
\begin{equation}
p(q_t\mid y_{1:t-1},\phi)
=
p(q_t\mid y_{1:t-1})
\label{eq:question_selection_independence}
\end{equation}
holds.
Furthermore, since the preference $\theta$ is time-invariant, the response at is conditionally independent of the past observation history $y_{1:t-1}$ given the question $q_t$ and $\phi$. That is,
\begin{equation}
p(a_t\mid q_t,y_{1:t-1},\phi)
=
p(a_t\mid q_t,\phi)
\label{eq:response_conditional_independence}
\end{equation}
holds.

\subsubsection*{Proof}
To simplify the notation, we define
\begin{equation}
p(y_1\mid y_{1:0},\phi)
:=
p(y_1\mid\phi).
\end{equation}
Then, the likelihood function of the full history of observations $y_{1:T}=(y_1,\dots,y_T)$ is expressed as
\begin{equation}
p(y_{1:T}\mid\phi)
=
\prod_{t=1}^{T}
p(y_t\mid y_{1:t-1},\phi).
\label{eq:sequential_likelihood}
\end{equation}
Therefore, the log-likelihood for the full history of observations is given by
\begin{align}
\log p(y_{1:T}\mid\phi)
&=
\sum_{t=1}^{T}
\log p(y_t\mid y_{1:t-1},\phi).
\label{eq:sequential_log_likelihood}
\end{align}
Differentiating \eqref{eq:sequential_log_likelihood} with respect to $\phi$ yields
\begin{equation}
\nabla_{\phi}\log p(y_{1:T}\mid\phi)
=
\sum_{t=1}^{T}
\nabla_{\phi}\log p(y_t\mid y_{1:t-1},\phi).
\label{eq:total_log_likelihood_gradient}
\end{equation}
Therefore, substituting  \eqref{eq:total_log_likelihood_gradient} into \eqref{total_fisher_information} yields
\begin{align}
I_{1:T}(\phi)
&= \mathbb{E}_{y_{1:T}\mid\phi}
\Biggl[
\left( \sum_{t=1}^{T} \nabla_{\phi}\log p(y_t\mid y_{1:t-1},\phi) \right) \nonumber \\
&\qquad\qquad  \left( \sum_{s=1}^{T} \nabla_{\phi}\log p(y_s\mid y_{1:s-1},\phi) \right)^{T}
\Biggr].
\label{eq:total_fisher_substitution}
\end{align}
Expanding \eqref{eq:total_fisher_substitution} yields
\begin{align}
I_{1:T}(\phi)
&= \sum_{t=1}^{T} \mathbb{E}_{y_{1:T}\mid\phi}
\Bigl[ \nabla_{\phi}\log p(y_t\mid y_{1:t-1},\phi) \nonumber \\
&\qquad\qquad\qquad \nabla_{\phi}\log p(y_t\mid y_{1:t-1},\phi)^{T} \Bigr] \nonumber \\
&\quad + \sum_{1\le t\neq s \le T} \mathbb{E}_{y_{1:T}\mid\phi}
\Bigl[ \nabla_{\phi}\log p(y_t\mid y_{1:t-1},\phi) \nonumber \\
&\qquad\qquad\qquad \nabla_{\phi}\log p(y_s\mid y_{1:s-1},\phi)^{T} \Bigr].
\label{eq:total_fisher_expansion}
\end{align}

We show that the cross-terms between different time steps appearing in the second term of  \eqref{eq:total_fisher_expansion} vanish.
To this end, we first show that
\begin{equation}
\mathbb{E}_{y_t\mid y_{1:t-1},\phi}
\left[
\nabla_{\phi}
\log p(y_t\mid y_{1:t-1},\phi)
\right]
=
0
\label{eq:conditional_gradient_zero}
\end{equation}
holds for each time $t$. 
First, from \eqref{conditional_gradient_reduction}, the following equation holds:
\begin{align}
&\mathbb{E}_{y_t\mid y_{1:t-1},\phi}
\left[ \nabla_{\phi} \log p(y_t\mid y_{1:t-1},\phi) \right] \nonumber \\
&= \sum_{a_t\in q_t} p(a_t\mid q_t,\phi) \nabla_{\phi} \log p(a_t\mid q_t,\phi).
\label{eq:conditional_gradient_expectation_1}
\end{align}
Applying the chain rule,
\begin{equation}
\nabla_{\phi}
\log p(a_t\mid q_t,\phi)
=
\frac{
\nabla_{\phi}p(a_t\mid q_t,\phi)
}{
p(a_t\mid q_t,\phi)
}
\label{eq:log_derivative_probability}
\end{equation}
holds.
Therefore,
\begin{align}
&\mathbb{E}_{y_t\mid y_{1:t-1},\phi}
\left[ \nabla_{\phi} \log p(y_t\mid y_{1:t-1},\phi) \right] \nonumber \\
&= \sum_{a_t\in q_t} p(a_t\mid q_t,\phi) \frac{ \nabla_{\phi}p(a_t\mid q_t,\phi) }{ p(a_t\mid q_t,\phi) } \nonumber \\
&= \sum_{a_t\in q_t} \nabla_{\phi} p(a_t\mid q_t,\phi) \nonumber \\
&= \nabla_{\phi} \sum_{a_t\in q_t} p(a_t\mid q_t,\phi) \nonumber \\
&= \nabla_{\phi}1 = 0.
\label{eq:conditional_gradient_expectation_2}
\end{align}
Thus, \eqref{eq:conditional_gradient_zero} is satisfied.
Next, using \eqref{eq:conditional_gradient_zero}, we show that the cross-terms between different time steps become 0.
First, let $s<t$. In this case, since $y_{1:s-1}$ is included in $y_{1:t-1}$, the expectation of $\nabla_{\phi}\log p(y_s\mid y_{1:s-1},\phi)$ is taken with respect to $y_{1:t-1}$. In addition, since the terms inside the expectation in \eqref{eq:total_fisher_expansion} do not include the observations from time $t+1$ to $T$, $y_{t+1:T}$, it suffices to take the expectation with respect to $y_{1:t}$. Furthermore, by Bayes' theorem, taking the expectation with respect to $y_{1:t}$ is equivalent to first taking the expectation with respect to $y_t\mid y_{1:t-1}$ and second taking the expectation with respect to $y_{1:t-1}$. 
Therefore, 
\begin{align}
&\mathbb{E}_{y_{1:T}\mid\phi}
\Bigl[ \nabla_{\phi} \log p(y_t\mid y_{1:t-1},\phi) \nonumber \\
&\qquad\qquad  \nabla_{\phi} \log p(y_s\mid y_{1:s-1},\phi)^{T} \Bigr] \nonumber \\
&= \mathbb{E}_{y_{1:t}\mid\phi}
\Bigl[ \nabla_{\phi} \log p(y_t\mid y_{1:t-1},\phi) \nonumber \\
&\qquad\qquad \nabla_{\phi} \log p(y_s\mid y_{1:s-1},\phi)^{T} \Bigr] \nonumber \\
&= \mathbb{E}_{y_{1:t-1}\mid\phi}
\Biggl[ \mathbb{E}_{y_t\mid y_{1:t-1},\phi} \bigl[ \nabla_{\phi} \log p(y_t\mid y_{1:t-1},\phi) \bigr] \nonumber \\
&\qquad\qquad\nabla_{\phi} \log p(y_s\mid y_{1:s-1},\phi)^{T} \Biggr] \nonumber \\
&= 0.
\label{eq:cross_term_s_less_t}
\end{align}
By symmetry, it also becomes 0 for $s>t$. 
From the above, for any case where $t\neq s$,
\begin{align}
&\mathbb{E}_{y_{1:T}\mid\phi}
\Bigl[ \nabla_{\phi} \log p(y_t\mid y_{1:t-1},\phi) \nonumber \\
&\qquad\qquad \nabla_{\phi} \log p(y_s\mid y_{1:s-1},\phi)^{T} \Bigr] = 0
\end{align}
holds.

Thus, since all the cross-terms between different time steps in \eqref{eq:total_fisher_expansion} vanish, we obtain
\begin{align}
I_{1:T}(\phi)
&= \sum_{t=1}^{T} \mathbb{E}_{y_{1:T}\mid\phi}
\Bigl[ \nabla_{\phi} \log p(y_t\mid y_{1:t-1},\phi) \nonumber \\
&\qquad\qquad  \nabla_{\phi} \log p(y_t\mid y_{1:t-1},\phi)^{T} \Bigr].
\label{eq:total_fisher_without_cross_terms}
\end{align}

Furthermore, by rewriting the terms inside the finite sum of \eqref{eq:total_fisher_without_cross_terms}, the following equation holds:
\begin{align}
&\mathbb{E}_{y_{1:T}\mid\phi}
\Bigl[ \nabla_{\phi} \log p(y_t\mid y_{1:t-1},\phi) \nonumber \\
&\qquad\qquad \nabla_{\phi} \log p(y_t\mid y_{1:t-1},\phi)^{T} \Bigr] \nonumber \\
&= \mathbb{E}_{y_{1:t-1}\mid\phi}
\Biggl[ \mathbb{E}_{y_t\mid y_{1:t-1},\phi} \Bigl[ \nabla_{\phi} \log p(y_t\mid y_{1:t-1},\phi) \nonumber \\
&\qquad\qquad  \nabla_{\phi} \log p(y_t\mid y_{1:t-1},\phi)^{T} \Bigr] \Biggr] \nonumber \\
&= \mathbb{E}_{y_{1:t-1}\mid\phi}
\left[ I_t(\phi\mid y_{1:t-1}) \right].
\label{eq:conditional_fisher_expectation}
\end{align}

Substituting \eqref{eq:conditional_fisher_expectation} into \eqref{eq:total_fisher_without_cross_terms} yields
\begin{equation}
I_{1:T}(\phi)
=
\sum_{t=1}^{T}
\mathbb{E}_{y_{1:t-1}\mid\phi}
\left[
I_t(\phi\mid y_{1:t-1})
\right],
\end{equation}
which completes the proof of Lemma 2. 

\begin{remark}
In the proof of Lemma 2, the most important point is that the cross-terms between different time steps vanish. This implies that the gradients of the log-likelihood at different time steps are orthogonal. To interpret this meaning more intuitively, it indicates that the direction for updating $\phi$ derived from the information of the observation $y_t$ at time $t$ does not depend on how $\phi$ was updated based on the past observations $y_{1:t-1}$.
\end{remark}

\section{Appendix: Proof of Lemma 3}
\subsubsection*{Proof}
The CDF of $X_i$ is calculated as follows:
\begin{align}
\label{eq:mtg06}
&F_i(x)\nonumber\\
&= \int_{-\infty}^{x} f_i(t)\,dt \nonumber \\
&= \int_{-\infty}^{x} \frac{1}{\beta}\exp\left(-\frac{t-\mu_0}{\beta}\right)  \exp\left(-\exp\left(-\frac{t-\mu_0}{\beta}\right)\right)dt.
\end{align}
By evaluating \eqref{eq:mtg06}, $F_i(x)$ can generally be written in the following explicit form:
\begin{align}\label{eq:mtg13}
F_i(x)=\exp\left(-\exp\left(-\frac{x-\mu_0}{\beta}\right)\right).
\end{align}

The probability $\gamma$ that the $\mathrm{EIG}^{\mathrm{eff}}$ value $X_1$ of $\mathcal{G}^{\ast}$ is greater than or equal to the $\mathrm{EIG}^{\mathrm{eff}}$ values $X_2, \dots, X_i, \dots, X_M$ of all the other $M-1$ blocks is expressed as
\begin{equation}
\label{eq:mtg14}
 \gamma=\int_{-\infty}^{\infty} f_1(x)\prod_{i=2}^{M}F_i(x)\,dx.
\end{equation}
Under the assumption that the $\mathrm{EIG}^{\mathrm{eff}}$ values of the $M-1$ groups are independent and identically distributed (i.i.d.), \eqref{eq:mtg14} is rewritten as
\begin{equation}
\label{eq:mtg15}
 \gamma=\int_{-\infty}^{\infty} f_1(x)\left(F_i(x)\right)^{M-1}dx.
\end{equation}
That is,
\begin{align}
\label{eq:mtg16}
\gamma
&= \int_{-\infty}^{\infty} \frac{1}{\beta}\exp\left(-\frac{x-\mu_1}{\beta}\right) \exp\left(-\exp\left(-\frac{x-\mu_1}{\beta}\right)\right) \nonumber \\
&\qquad \qquad \cdot \left( \exp\left(-\exp\left(-\frac{x-\mu_0}{\beta}\right)\right) \right)^{M-1}dx.
\end{align}
Let
\begin{equation}
\label{eq:mtg17}
 A_1=e^{\mu_1/\beta},\qquad A_0=e^{\mu_0/\beta},
\end{equation}
\eqref{eq:mtg16} is written as
\begin{equation}
\label{eq:mtg18}
 \gamma=\int_{-\infty}^{\infty}\frac{1}{\beta}A_1 e^{-x/\beta} \exp\left(-(A_1+(M-1)A_0)e^{-x/\beta}\right)dx.
\end{equation}
Furthermore, introducing the variable change
\begin{equation}
\label{eq:mtg19}
 z=e^{-x/\beta},
\end{equation}
we obtain
\begin{equation}
\label{eq:mtg20}
 dz=-\frac{1}{\beta}e^{-x/\beta}dx.
\end{equation}
As $x$ goes from $-\infty$ to $\infty$, $z$ goes from $\infty$ to $0$. Therefore, by defining $d:=\frac{\mu_1-\mu_0}{\beta}$, \eqref{eq:mtg18} is computed as
\begin{align}
\gamma
&=\int_{\infty}^{0}A_1\exp\left(-(A_1+(M-1)A_0)z\right)(-dz) \nonumber\\
&=\int_{0}^{\infty}A_1\exp\left(-(A_1+(M-1)A_0)z\right)dz \nonumber\\
&=A_1\left[\frac{\exp(-(A_1+(M-1)A_0)z)}{-(A_1+(M-1)A_0)}\right]_{0}^{\infty} \nonumber\\
&=\frac{A_1}{A_1+(M-1)A_0} \nonumber\\
&=\frac{e^d}{e^d+M-1}.
\label{eq:mtg21}
\end{align}
\eqref{eq:mtg21} completes the proof of Lemma 3. 

\section{Appendix: Validity of Assumption 2}
We empirically verified the validity of assuming that the $\mathrm{EIG}^{\mathrm{eff}}$ values follow a Gumbel distribution through a numerical experiment. Specifically, we conducted a preference estimation simulation and collected the $\mathrm{EIG}^{\mathrm{eff}}$ values at each time step, separating them into two categories: values for questions included in $\mathcal{G}^{\ast}$ and values for the remaining questions. Using these separated datasets, we constructed two normalized histograms. We then fitted the PDF of the Gumbel distribution to each of these histograms. By evaluating the goodness of fit between the estimated Gumbel distributions and the normalized histograms, we verified the validity of the Gumbel distribution assumption for the $\mathrm{EIG}^{\mathrm{eff}}$ values. The simulation run covered up to 100 time steps. 

\begin{figure}[t]
  \centering
  \includegraphics[width=1\linewidth]{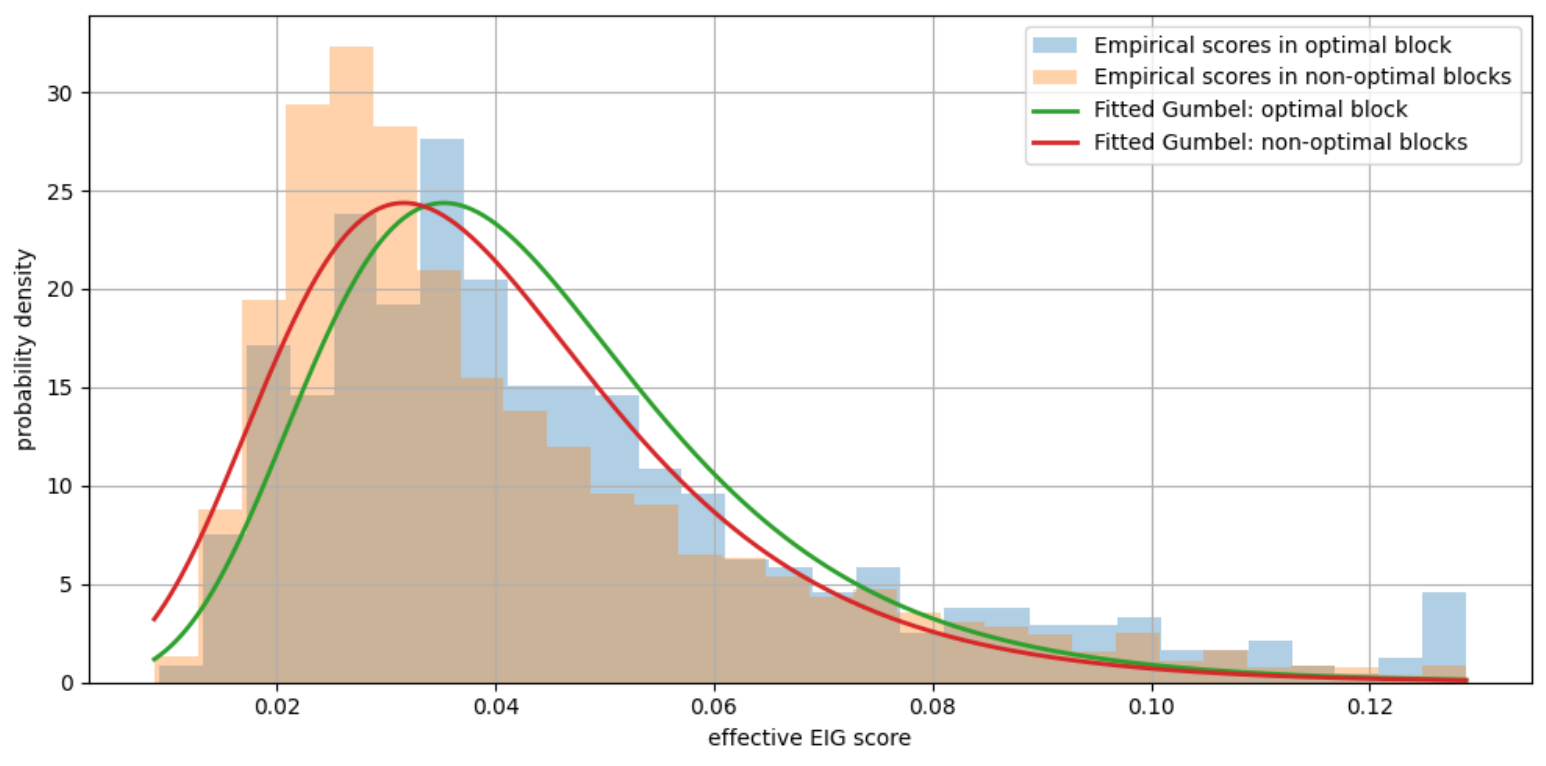}
  \caption{Histograms of $\mathrm{EIG}^{\mathrm{eff}}$ values across all time steps and the fitted Gumbel distributions.}
  \label{Gumbel}
\end{figure}

The results are shown in Fig.~\ref{Gumbel}.
As can be seen from Fig.~\ref{Gumbel}, the shapes of the $\mathrm{EIG}^{\mathrm{eff}}$ value histograms and the fitted Gumbel distributions are remarkably similar. These results suggest that assuming a Gumbel distribution for the $\mathrm{EIG}^{\mathrm{eff}}$ is generally reasonable.


\section*{Acknowledgment}
We would like to thank Yuki Miyoshi , belonging to Keio University, for useful discussions.

                                        
\end{document}